\documentclass[
    a4paper,
    man,
    british
]{apa6}

\usepackage[british]{babel}
\usepackage[utf8]{inputenc}
\usepackage{epstopdf}
\usepackage{csquotes}
\usepackage[hidelinks]{hyperref}
\usepackage[
    style=apa,
    backend=biber,
    sortcites=true,
    sorting=nyt,
    hyperref=false,
    backref=false,
]{biblatex}

\DeclareLanguageMapping{british}{british-apa}

\let \cite \parencite

\renewbibmacro*{related:reprintfrom}[1]{%
  \entrydata*{#1}{%
    \printtext{\mkbibemph{\printfield[apacase]{title}}}%
    \setunit{\bibpagespunct}%
    \printfield{pages}%
    \setunit{\addcomma\addspace}%
    \bibstring{byauthor}\addspace
    \printnames[apanames][-\value{listtotal}]{editor}%
    \ifnameundef{editor}
      {}
      {\addcomma\addspace
       \usebibmacro{apaeditorstrg}{editor}}
    \printnames[apanames][-\value{listtotal}]{author}%
    \setunit{\addcomma\addspace}%
    \usebibmacro{date}%
    \setunit{\addcomma\addspace}%
    \usebibmacro{location+publisher}%
    \newunit\newblock
    \usebibmacro{related}}}

\DefineBibliographyStrings{british}{
  reprintfrom = {Reprinted from}
}

\bibliography{bibliography}

\title{The Scientific Evidence Indicator for Popular Science News}
\shorttitle{The Scientific Evidence Indicator}
\author{Anders Sundnes Løvlie, Astrid Waagstein and Peter Hyldgård}
\affiliation{IT University of Copenhagen and videnskab.dk}

\keywords{design, science communication, health, post-truth, hci}

\begin{document}

\maketitle

\section{Introduction}
According to McIntyre, one of the roots of the so-called "post-truth" phenomenon \cite{mcintyre_post-truth_2018} is "science denialism", which is characterized by media reports that suggest uncertainty or controversy about issues where there in fact is consensus in the scientific community. This presents an inaccurate image of the state of scientific debate, making it harder to fight problems like tobacco, global warming or opposition to vaccination. The World Health Organization has named "vaccine hesitancy" one of the top threats to global health in 2019, suggesting more "trusted, credible information" as the main remedy \cite{who}. However, the fact that many people continue to believe falsehoods that scientists have long rejected, demonstrates that it is challenging for many to fully understand or trust the ways in which scientific truths are determined.

To what extent can news media help in providing more credible information about science? This is the core challenge for the Science Evidence Indicator (SEI) project, a collaboration between the Danish popular news website videnskab.dk and the authors of this paper. Looking specifically at medical science news, we aim to provide a transparent assessment of the scientific sources behind a story. This entails identifying some of the criteria that scientists use to assess research, and making it accessible and understandable for readers. We address the following research question: \textit{How can we communicate the quality of scientific publications in health science to a non-expert audience?} Our goal is to make the assessments understandable for the youngest part of the website's target audience: high school students from age 16 and upwards.

\section{Related work}
Research in science communication has suggested a need for improved metrics for science \cite{Treise2002-ac,Weigold2001-pl}, and for communicating about scientific uncertainty and credibility \cite{National_Academies_of_Sciences_Engineering_Medicine2017-eh, Berdahl2016-zi, Weingart2016-om, Retzbach2015-os}. Research on science communication and online health information emphasizes the importance of visual design \cite{Eastin2006-cj,Metzger2007-gm, Tufte2001-bb, Pauwels2006-bq, Allen2018-oq, Li2018-ez,Tal2016-me}. Takahashi and Tandoc suggest that lack of trust in the news media may work as an incitement to learn about science \cite{Takahashi2016-ga, Tandoc2018-wz}. A survey of health journalists reveal a tension between simplifying technical language to improve comprehension, and maintaining scientific credibility \cite{Hinnant2009-op}. Oxman et al.'s "Index of Scientific Quality" for news about medical research relies on expert input and would therefore be costly to apply \cite{Oxman1993-wi}. Viviani and Pasi point to a lack of research on automatic credibility assessment in this area \cite{Viviani2017-mo}.

\section{Method}
The Science Evidence Indicator has been developed as a research through design \cite{zimmerman_research_2007} project in collaboration with representatives of the videnskab.dk website. The design has been developed through an iterative process and has undergone a series of small-scale, qualitative user tests.
The SEI is being implemented in the website's publishing system to be launched "in the wild" during the spring of 2019.

\begin{figure}
  \includegraphics{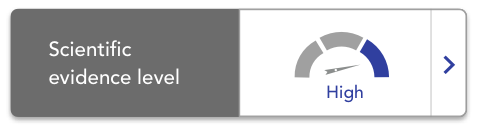}
  \caption{The Scientific Evidence Indicator.}
  \label{fig1}
\end{figure}

\section{Design}

\begin{figure}
  \centering
  \includegraphics[scale=0.3]{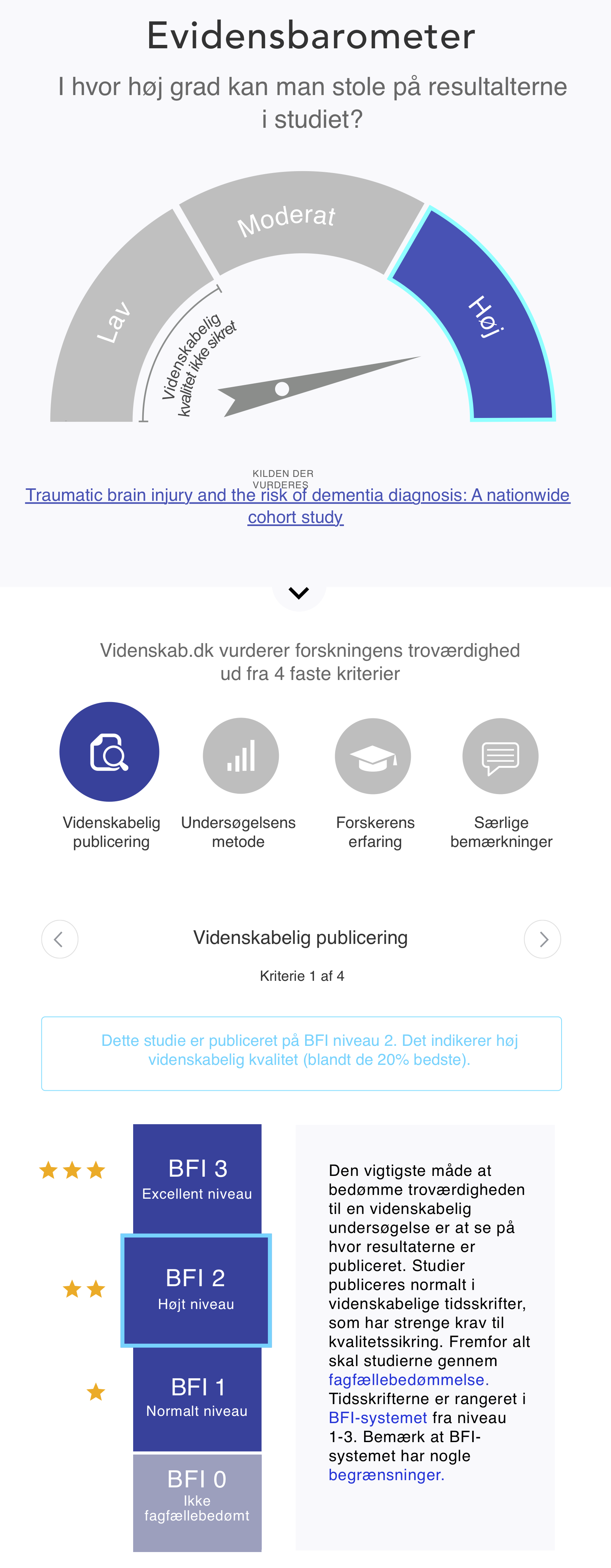}
  \caption{The expanded Scientific Evidence Indicator (in Danish).}
  \label{fig2}
\end{figure}

The SEI is designed as a module to be added to news stories about new research in the medical sciences at videnskab.dk. It consists of a small graphic placed at the top of the news story, indicating the "scientific evidence level" for the sources of the article (Fig. \ref{fig1}). The graphic expands to a larger version which explains the assessment and links to further material explaining how scientists assess evidence, and the principles and limitations of the SEI (Fig. \ref{fig2}). The assessment is based on four separate quality indicators:

\begin{enumerate}
\item \textit{Scientific publication}: This variable uses the Danish Bibliometric Research Indicator (BFI) to gauge whether the study has been published in a peer reviewed, meritable scientific publication channel. It gives the number 0 if the publication does not meet the BFI's minimum standard, and otherwise 1-3 depending on the BFI score of the publication. (For information about BFI, see \cite{danish_agency_for_science_and_higher_education_guidelines_2019}.)
\item \textit{Method}: This variable uses a 7-point model of the evidence hierarchy for medical research, to indicate the strength of the conclusions in the study. 
\item  \textit{Researcher's Experience}: This variable uses the H-index of the highest ranked among the authors of the study as a measure of the "experience" of the scientific team. The scale has four levels: "Excellent" (60+), "Very Experienced" (40-60), "Experienced" (20-40) and "Less Experienced" (0-20).
\item  \textit{Special Remarks}: Used by journalists to point out important aspects of a study that is relevant when assessing the reliability of the conclusions. Journalists are required to add an explanation if they are reporting on a study that has not been peer reviewed according to the BFI standard.
\end{enumerate}

The variables in the SEI are filled in manually by the journalist when adding a scientific source to a news story, resulting in an aggregate score for that source's "Scientific Evidence Level":
\begin{itemize}
\item Low - if the source is not in the BFI system.
\item Medium - if the source has been published at BFI level 1 or higher, but does not meet all the criteria for the 'high' score.
\item High - if the source is published at BFI level 2 or 3, the method is ranked as one of the two top levels of the evidence hierarchy, and the h-index of at least one of the authors is above 20.
\end{itemize}
While the criterion for 'low' rests on the authority of the BFI system, the criteria for distinguishing between 'medium' and 'high' do not reflect any kind of scholarly consensus, but is rather a heuristic developed in order to identify sources that have exceptionally high level of reliability.

\section{Evaluation}
Tests of early prototype versions demonstrated a strong need to simplify technical terms, and to provide an aggregate score for each study, while balancing the need for nuance and accuracy. The name of the indicator as well as the variables used, and their explanations, have been revised repeatedly in order to improve comprehension.


In order to test whether the final design presented here is comprehensible to readers in the target audience, we arranged a test in a high school class consisting of 14 students aged 16-17, in October 2018. The students were presented with three recent stories about the health effects of cannabis in a clickable prototype for the SEI and whose scientific sources were assessed as low, medium and high. After reading the three articles the students were asked to answer an online questionnaire and participate in a short group interview. Based on a preliminary analysis of these data we find that the SEI seems to have been effective at \textit{informing} the students about the scientific evidence level for each of the three articles, but that there are challenges with providing a more fundamental \textit{understanding} of the SEI assessment. Nearly all the students were able to correctly state the "evidence level" of each study (according to the SEI scale) in the online questionnaire. However, only around one third of the students were able to correctly identify that the assessment was based on the four variables presented above, and most struggled to explain in their own words the meaning of these variables. This is not surprising, since this is a complicated matter that many university students also struggle to understand. Given that the SEI is intended to be a permanent part of the website, we hope that regular readers may gradually understand the indicator better as they encounter it on different stories.

\section{Further research}
When the SEI is implemented on the videnskab.dk website we plan to set up an A/B test coupled with an online survey in order to assess the degree to which the indicator is understandable to a broad audience. Furthermore, videnskab.dk delivers content to many other media partners, and are interested in developing the SEI further towards a system that can be applied by a broader range of media organizations, both in Denmark and abroad.

Dependent on more funding, we are also interested in developing the system further to a partially or wholly automatized tool in which the three variables can be scored using information in online publication databases such as Google Scholar and PubMed. This would enable us to expand the scope of the tool beyond articles written by journalists - e.g. allowing users to get an automatized assessment of any medical science publication. The purpose would be not to create a ranking system for publications, but rather to help users separate "pseudoscience" from reliable science. 

\begin{figure}

  \includegraphics[width=\linewidth]{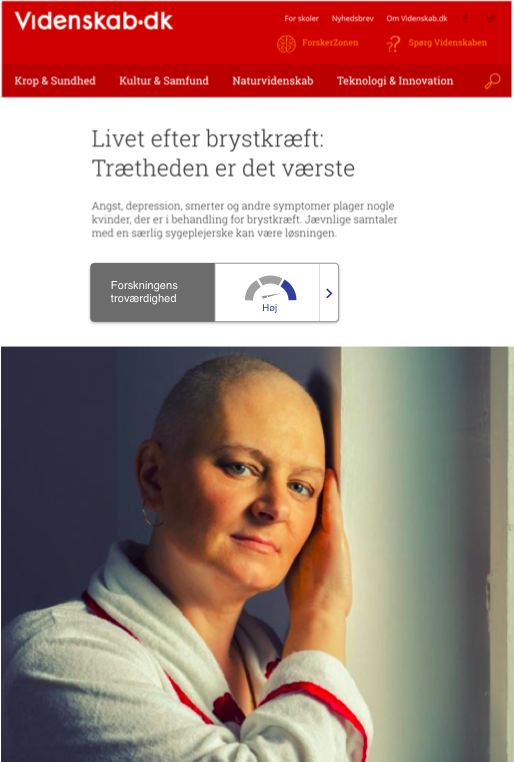}
  \caption{News story from videnskab.dk with the SEI placed just after the lead.}
  \label{fig_placering}
\end{figure}

\section{Acknowledgments}
This project has received funding from Google's Digital News Innovation Fund. The following master students made valuable contributions to the design presented here: S\o ren Gollander-Jensen, Louis Valman H\o ffding Dyrhauge, Anders Steen Mikkelsen and Martin Rust Priis Christensen.

\printbibliography

\end{document}